\documentclass[conference,10pt]{IEEEtran}

\usepackage{cite}
\ifCLASSINFOpdf
   \usepackage[pdftex]{graphicx}
\else
  \usepackage[dvips]{graphicx}
\fi

\usepackage[cmex10]{amsmath}
\usepackage{amssymb}
\usepackage{amsfonts}
\usepackage{amsthm}
\usepackage{array}
\usepackage{dsfont}
\usepackage{setspace}
\usepackage{epstopdf}
\usepackage[english]{babel}
\usepackage{color}
\usepackage{bbm}

\newtheorem{lemma}{Lemma}

\newtheorem{corollary}{Corollary}
\newtheorem{theorem}{Theorem}
\newtheorem{proposition}{Proposition}

\long\def\symbolfootnote[#1]#2{\begingroup%
\def\thefootnote{\fnsymbol{footnote}}\footnote[#1]{#2}\endgroup}
\allowdisplaybreaks[1]

\IEEEoverridecommandlockouts

\begin{document}

\title{Smart Meter Privacy with Renewable Energy \\ and a Finite Capacity Battery}

\author{
  \IEEEauthorblockN{Giulio Giaconi and Deniz G\"{u}nd\"{u}z}

  \IEEEauthorblockA{Department of Electrical and Electronic Engineering, Imperial College London,  London, UK\\
    {\{g.giaconi, d.gunduz\}}{@imperial.ac.uk}
}
\thanks{This work is supported in part by the UK Engineering and Physical Sciences Research Council (EPSRC) under grant number EP/N021738/1.}


}

\maketitle

\begin{abstract}
We address the \emph{smart meter} (SM) privacy problem by considering the availability of a renewable energy source (RES) and a battery which can be exploited by a consumer to partially hide the  consumption pattern from the utility provider (UP). Privacy is measured by the mutual information rate between the consumer's energy consumption and the renewable energy generation process, and the energy received from the grid, where the latter is known by the UP through the SM readings, and the former two are to be kept private. By expressing the information leakage as an additive quantity, we cast the problem as a stochastic control problem, and formulate the corresponding Bellman equations.
\end{abstract}


\section{Introduction}

An essential component of a smart grid is the \emph{smart meter} (SM), a device that records minutely the electricity consumption of a household. The adoption of SMs is a key advantage for both utility providers (UPs), who would be able to better monitor the consumption and trade energy with users, and the distribution system operators, who would better manage and run the network. The adoption of SMs is also favourable for consumers, since it allows a time-of-usage pricing with the consequent possibility to reduce electricity costs by choosing less expensive time slots for power-consuming appliances.

However, the high resolution of the data collected by the SMs also makes it possible to infer a consumer's energy consumption \emph{load profile}, i.e., the time series of energy usage collected with regularity from a household. These profiles are extremely valuable, since it is possible to extrapolate sensitive information from them, such as users' habits and presence at home, illnesses or disabilities, the equipments being used, and even which TV channel is being watched \cite{Greveler:2012}.

Approaches to address SM privacy in the literature can be broadly classified into two groups: those that modify the SM readings before being sent to the UP, and those that modify the actual user energy demand. In the first group, obfuscation \cite{Kim:2011}, anonymization \cite{Efthymiou:2010} and aggregation techniques \cite{Bohli:2010} are included. Regarding the second approach, user consumption can be filtered through a storage device, as described in \cite{Kalogridis:2010}, \cite{Koo:2012}, \cite{Varodayan:2011}, \cite{Zuxing:2016ICASSP} and \cite{Li:2015}, or also by including an alternative energy source, e.g., a renewable energy source (RES), as in \cite{Tan:2013}, \cite{Gomez:2013ISIT} and \cite{Giaconi:2015}. In particular, in \cite{Zuxing:2016ICASSP}, privacy is evaluated through a Bayesian detection setting and the problem is formulated as a Markov decision process (MDP). The smart grid state estimation problem is addressed in \cite{Sandberg:2015} and the trade-off between differential privacy and the mean distortion of the state estimate is studied. Despite all these efforts, SM privacy is an area of ongoing active research, and wide consensus is still to be reached even on fundamental questions, such as how to measure privacy.

In this paper, we study the SM privacy problem in the presence of a RES together with a rechargeable energy storage device, i.e., a battery. We adopt an information theoretic approach, by minimizing the mutual information between the input load and the renewable energy process, and the output load of the system. Our main contribution here is to cast this problem as an MDP, by finding an additive formulation for the information leakage. We note that a similar approach has been followed in \cite{Li:2015}, where a battery has been considered. Here, we also consider the presence of a RES to further hide user's consumption. Finally, the corresponding dynamic program (DP) is formulated, which can be solved numerically in order to identify the optimal energy management policy.

The remainder of this paper is organized as follows. In Section \ref{sec:SystemModel} the system model is introduced. In Section \ref{sec:EnergyNotKnown} it is assumed that the UP does not know the realizations of the renewable energy process, whereas in Section \ref{sec:EnergyKnown} it does. For both cases, the information leakage minimization problem can be cast as an MDP. Conclusions are drawn in Section \ref{sec:conclusion}.


\section{System Model}\label{sec:SystemModel}
\begin{figure}[!t]
\centering
\includegraphics[width=1\columnwidth]{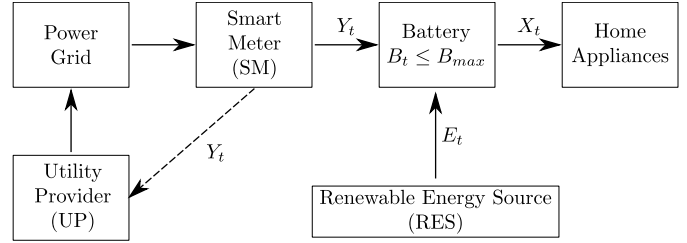}
\caption{System model. $X_t$, $Y_t$, $E_t$ and $B_t$ are the user energy demand, the output load, the amount of renewable energy obtained from a renewable energy source, and the state of the battery at time $t$, respectively. The dashed line highlights the meter readings being reported to the UP.}
\label{fig:SystemModel}
\end{figure}
We consider the discrete time system depicted in Figure  \ref{fig:SystemModel}. The input load $X_t \in \mathcal{X}$ is the energy requested by the user in time slot $t$, where $\mathcal{X}=\{0,1,\ldots X_{\max}\}$. The output load $Y_t \in \mathcal{Y}$ is the energy received from the UP, where $\mathcal{Y}=\{0,1,\ldots,Y_{\max}\}$. We assume $\{X_t\}_{t=1}^{\infty}$ to be a first-order time-homogeneous Markov chain with transition probability $q_X$, whose initial state $X_1$ is distributed according to $p_{X_{1}}$. In time slot $t$, $E_t\in \mathcal{E} = \{0,1, \ldots E_{\max}\}$ units of energy are generated by the RES, which becomes available at the beginning of time slot $t$. $\{E_t\}_{t=1}^{\infty}$ is also a first-order time-homogeneous Markov chain with transition probability $q_E$ and initial state $E_1$, distributed according to $p_{E_{1}}$. We further assume the availability of a battery of capacity $B_{\max}$, whose state of charge at the beginning of time slot $t$ is $B_t\in \{0,1,\ldots B_{\max}\}$. The initial state of the battery $B_1$ has distribution $p_{B_1}$. We assume that all the involved random processes are defined over finite alphabets and that there is a minimum unit of energy such that all the aforementioned quantities are integer multiples of this unit. Moreover, we assume $Y_{\max}\geq X_{\max}+B_{\max}$.

In our model, user demands have to be satisfied at all times:
\begin{equation}\label{eq:yConstraint2}
B_t+E_t+Y_t \geq X_t, \qquad \forall t.
\end{equation}

We do not allow intentional energy waste, or selling energy back to the grid; that is, we impose
\begin{equation}\label{eq:yConstraint3}
0 \leq Y_t \leq \big[B_{\max} + X_t - B_t -E_t \big]^{+}, \qquad \forall t,
\end{equation}
where $[a]^{+} = a$ if $a>0$, and $0$ otherwise.
Similarly, we do not allow wasting of renewable energy when the battery is not full. These actions could provide additional privacy to the user, albeit at a significantly higher energy cost. However, energy can still be wasted inevitably, for example, when the battery is full and the input load is smaller than the generated energy. The battery state is updated as
\begin{equation}\label{eq:battery_constraint}
	B_{t+1} = \min\Big\{B_{t} + E_t - X_t,B_{\max}\Big\}+Y_t, \qquad \forall t.
\end{equation}


Let $W_t \triangleq B_t + E_t -X_t$. The feasible set for $Y_t$, given $W_t= w_t$, is given by
\begin{equation*}
\mathcal{\bar{Y}}(w_t) \triangleq \Big\{ y_t \in \mathcal{Y} : [-w_t]^{+} \leq y_t \leq \big[B_{\max} - w_t \big]^{+} \Big\}.
\end{equation*}


An energy management policy $f=(f_1,f_2,\ldots)$ decides on the amount of energy to request from the UP at each time $t$, given the previous values of input load $X^t$, renewable energy $E^{t}$, battery state $B^{t}$, and output load $Y^{t-1}$. We consider randomized policies, that is, each $f_t$ is a conditional probability $f_t(y_t|x^t,e^{t},b^t,y^{t-1})$, with $f \in \mathcal{F}$.

Our goal is to minimize the leakage to the UP of information about user's energy consumption as well as the amount of energy generated by the RES. Accordingly, the information leakage rate induced by policy $f$ over $n$ time slots is
\begin{equation}\label{eq:privacyFunction}
\mathcal{I}(n,f) \triangleq \frac{1}{n} I^{f}(X^n, E^n,B_1;Y^n).
\end{equation}

Thus, our optimization problem can be written as
\begin{equation}\label{eq:generalMinimization}
\mathcal{I}(n) \triangleq \inf_{f \in \mathcal{F}} \frac{1}{n} I^{f}(X^n, E^n,B_1;Y^n),
\end{equation}
where the minimization is over all the feasible policies $f \in \mathcal{F}$. Note that, in the above form, this is an extremely complex optimization problem as we need to specify $p(Y_t|X^{t}, E^{t}, B_1, Y^{t-1})$ for every $t$, and every possible realization $(X^{t}, E^{t}, B_1, Y^{t-1}) = (x^{t}, e^{t}, b_1, y^{t-1})$.

In the following, we identify two different scenarios based on the information available at the UP regarding the renewable energy process. For both scenarios, we will rewrite (\ref{eq:generalMinimization}) in an additive form, which consequently can be formulated as a DP. Given $q_X$, $q_E$, $p_{X_1}$, $p_{E_1}$ and $p_{B_1}$, our goal is to find an optimal battery charging policy $f=(f_1,f_2,\ldots) \in \mathcal{F}$ that solves (\ref{eq:generalMinimization}), for both finite and infinite horizon settings.





\section{Renewable Energy not Known by the UP}\label{sec:EnergyNotKnown}

In this section $E^n$ is treated as a random sequence whose realizations are known causally only to the user. This scenario may occur if, for example, $E^n$ originates from light, vibration, thermal, or biological sources, which could be extremely difficult, if not impossible, for the UP to track.

\subsection{Additive Formulation for the Information Leakage Rate}

To formulate the problem as an MDP we need to write the cost function, i.e., the leakage rate, in an additive form. Specifically, we aim at a formulation in which, conditioned on $Y^{t-1}$, the output load at time t, $Y_t$, leaks information only on $X_t$, $E_t$ and $B_t$, but not on their past values. This will be achieved by restricting the set of possible policies to a smaller set $\mathcal{F}' \subseteq \mathcal{F}$, where each $f' \in \mathcal{F}'$ is a conditional probability of the form $f'_t(y_t|x_t,e_{t},b_t,y^{t-1})$, i.e.,
\begin{equation*}
f'_t: \mathcal{X} \times \mathcal{E} \times \mathcal{B} \times \mathcal{Y}^{t-1} \rightarrow \mathcal{Y}, \qquad \forall t.
\end{equation*}

The joint measure induced by $f'$ is
\begin{IEEEeqnarray}{rcl}\label{eq:pdf2}
  &p^{f'}&(X^n=x^n,E^n=e^n,B^n=b^n,Y^n=y^n) \nonumber\\
  &=& p_{X_1}(x_1)p_{E_1}(e_1)p_{B_1}(b_1)f'_{1}(y_1|x_1,e_1,b_1) \nonumber \\
  && \times\prod_{t=2}^{n}\bigg[\mathbbm{1}_{b_t} \Big\{\min\{b_{t-1} + e_{t-1} - x_{t-1},B_{\max}\}+y_{t-1}\Big\} \nonumber \\
  && \times q_X(x_t|x_{t-1})q_E(e_t|e_{t-1}) f'_t(y_t|x_t,e_t,b_t,y^{t-1})\bigg], \nonumber
\end{IEEEeqnarray}
where $\mathbbm{1}$ is the indicator function, i.e., $\mathbbm{1}_{b_t}\{a\} = 1$ if $b_t= a$, and $0$ otherwise. $b_t \in \{0, \ldots, B_{\max}\}$ holds since $Y_t \in \bar{\mathcal{Y}}(w_t)$.

The next theorem, whose proof is omitted due to space limitations, states that this restriction is without loss of optimality.

\begin{theorem}\label{th:theoremppf}
There is no loss of optimality in focusing only on charging strategies $f' \in \mathcal{F}'$, where $\mathcal{F}'\subseteq \mathcal{F}$. Moreover, the minimum information leakage rate can be written in the following additive form
\begin{equation}\label{eq:theorem1ppf}
 \mathcal{I}(n) = \inf_{f' \in \mathcal{F}'} \frac{1}{n}  \sum_{t=1}^{n} I^{f'}(X_t,E_t,B_t;Y_t|Y^{t-1}).
\end{equation}
\end{theorem}


It is possible to generalize Theorem \ref{th:theoremppf} to the scenario in which $X$ and $E$ are Markov chains with order higher than $1$, i.e., $p_{X_t|X^{t-1}}=p_{X_t|X^{t-1}_{t-m}}$ and $p_{E_t|E^{t-1}}=p_{E_t|E_{t-l}^{t-1}}$. If new policies $f'' \in \mathcal{F}''$ are defined such that
\begin{equation}\label{eq:policyGreaterOrder}
f''_t:\mathcal{X}^m \times \mathcal{E}^{l} \times \mathcal{B} \times \mathcal{Y}^{t-1} \rightarrow \mathcal{Y}, \qquad \forall t,
\end{equation}
where $\mathcal{F}'' \subseteq \mathcal{F}$, then the following corollary holds.
\begin{corollary}\label{th:greaterOrder}
Let $X$ and $E$ be Markov chains of order $m$ and $l$, respectively. There is no loss of optimality in focusing only on charging strategies $f'' \in \mathcal{F}''$, and for those strategies the minimum information leakage rate can be written in the following additive form
\begin{equation*}
 \mathcal{I}(n) = \inf_{f'' \in \mathcal{F}''} \frac{1}{n}  \sum_{t=1}^{n} I^{f''}(X_{t-m+1}^t,E_{t-l+1}^t,B_t;Y_t|Y^{t-1}).
\end{equation*}

\end{corollary}


\subsection{MDP Formulation}\label{subsec:MDP}

Our next goal is to cast the problem as a stochastic control problem, which can be formulated as a DP. For this, we need to specify the state space, the control actions and the instantaneous cost corresponding to state-action pairs. The per-step cost in (\ref{eq:theorem1ppf}) depends on past observations $y^{t-1}$, which could be considered as the state of the DP at time $t$. However, this would mean a state space growing with time. To avoid this, we follow the approach of \cite{Tatikonda:2009} and \cite{Li:2015}, and introduce a belief state, which can be shown to replace the $y^{t-1}$ sequence.

The state of the DP at time $t$ is considered to be the causal posterior probability distribution over the triplet $S_t \triangleq (X_t,E_t,B_t)$, given the knowledge of past outputs $Y^{t-1}$:
\begin{equation}\label{eq:state}
\beta_t(s_t) \triangleq
\begin{cases}
p(s_1), & \text{if } t=1, \\
p(s_t|y^{t-1}),       & \text{otherwise}.
\end{cases}
\end{equation}
$\beta_t(s_t)$ can be considered as the \emph{belief} that the UP has about $s_t$ at time $t$, given its past observations $Y^{t-1}$.

The control action $U_t \in \mathcal{U}$ is the conditional probability $u_t(y_t|s_t)$. A \emph{randomized history-dependent policy} $\pi=\{\pi_1, \pi_2, \ldots\}$ chooses control action at time $t$ via $u_t=\pi_t(h)$, where $h$ represents the history available to the controller. Thus, the time ordering of the events is $S_1, U_1, Y_1, S_2, U_2, Y_2, \ldots$.

Without loss of optimality, we can focus on randomized \emph{Markov policies} that depend only on the current state $\beta_t$, i.e., $\pi_t(\beta_t)=\pi_t(h)$. This holds for both finite and infinite horizon problems under mild assumptions \cite{Bertsekas:2007}. Policy $\pi$ induces the following joint measure:
\begin{IEEEeqnarray*}{rcl}
  &p^{\pi}&(X^n=x^n,E^n=e^n,B^n=b^n,Y^n=y^n) \nonumber\\
  &=& p_{X_1}(x_1)p_{E_1}(e_1)p_{B_1}(b_1)u_1(y_1|x_1,e_1,b_1)    \nonumber  \\
  && \times \prod_{t=2}^{n}\bigg[\mathbbm{1}_{b_t} \Big\{\min\{b_{t-1} + e_{t-1} - x_{t-1},B_{\max}\}+y_{t-1}\Big\} \nonumber \\
  && \times q_X(x_t|x_{t-1})q_E(e_t|e_{t-1})u_t(y_t|x_t,e_t,b_t)\bigg].
\end{IEEEeqnarray*}

The state can be updated recursively, i.e., $\beta_{t+1}=\phi(\beta_t,u_t)$
\begin{IEEEeqnarray}{rcl}
&\beta_{t+1}&(s_{t+1}) = p(s_{t+1}|y^{t}),\nonumber \\
&=& \sum_{s_{t}} p(s_t,s_{t+1}|y^{t}),\nonumber\\
&=& \sum_{s_{t}} \frac{p(s_t,s_{t+1},y_{t}|y^{t-1})}{p(y_{t}|y^{t-1})},\nonumber\\
&\stackrel{(a)}{=}& \frac{\sum_{s_{t}} p(s_{t}|y^{t-1}) p(y_{t}|s_{t},y^{t-1}) p(s_{t+1}|s_{t},y_{t})}{\sum_{s_{t},s_{t+1}} p(s_{t}|y^{t-1}) p(y_{t}|s_{t},y^{t-1}) p(s_{t+1}|s_{t},y_{t}) },\nonumber\\
&\stackrel{(b)}{=}& \frac{\sum_{s_{t}} \beta_{t}(s_{t}) u_{t}(y_{t}|s_{t}) q_X(x_{t+1}|x_{t}) q_E(e_{t+1}|e_{t})}{\sum_{s_{t},s_{t+1}} \beta_{t}(s_{t}) u_{t}(y_{t}|s_{t}) q_X(x_{t+1}|x_{t}) q_E(e_{t+1}|e_{t})}\nonumber\\
&& \times \frac{ \mathbbm{1}_{b_{t+1}}\Big\{\min\{b_{t}+e_{t}-x_{t},B_{\max}\}+y_t\Big\}} {\mathbbm{1}_{b_{t+1}}\Big\{\min\{b_{t}+e_{t}-x_{t},B_{\max}\}+y_t\Big\}}, \label{eq:recursion}
\end{IEEEeqnarray}
where $(a)$ follows from Bayes rule and the Markov chain $Y^{t-1} \rightarrow (S_{t},Y_{t}) \rightarrow S_{t+1}$; and $(b)$ is due to the definitions of $\beta_t$ and $u_t$.


Given $Y^{t-1}=y^{t-1}$, the per-step cost of taking action $u_t$ when $s_t=(x_t,e_t,b_t)$ is
\begin{equation}\label{eq:instCost}
g_t(x_t,e_t,b_t,u_t,y^{t}) \triangleq \log \frac{u_t(y_t|x_t,e_t,b_t)}{p(y_t|y^{t-1})}.
\end{equation}

It is possible to show that this new formulation is equivalent to the original problem, by considering the average $n$-horizon cost and the knowledge of $y^{t-1}$ as in $\mathcal{I}(n,f')$
\begin{IEEEeqnarray}{rCl}
  \mathcal{I}(n,\pi) &=& \frac{1}{n} \mathbb{E}^{\pi} \bigg[ \sum_{t=1}^{n}  g_t(x_t,e_t,b_t,u_t,y^t)\bigg], \nonumber \\
  &=& \frac{1}{n} \sum_{t=1}^{n} \sum_{s_t\in \mathcal{S}, y^t \in \mathcal{Y}^t} p(s_t,y_t|y^{t-1}) \log \frac{u_t(y_t|s_t)}{p(y_t|y^{t-1})}, \nonumber\\
  &=& \mathcal{I}(n,f'),
\end{IEEEeqnarray}
where we remind that $u_t$ is also a function of $y^{t-1}$, since $u_t=\pi_t(\beta_t)$.
Given a policy $\pi$, $\beta_t$ and $u_t$ are determined by $y^{t-1}$. The average information leakage at time $t$ is
\begin{IEEEeqnarray}{rcl}\label{eq:costUpdate}
\mathbb{E}^{\mathbf{\pi}}\Big[&g_t&(X_t,E_t,B_t,U_t,Y^t)\Big] \nonumber \\
&=& I(X_t, E_t, B_t;Y_t|Y^{t-1}=y^{t-1}),\nonumber \\
&=& \sum_{\substack{x_t \in \mathcal{X}, e_t \in \mathcal{E} \\ b_t \in \mathcal{B}, y^t \in \mathcal{Y}^t}}  p(x_t,e_t,b_t|y^{t-1}) p(y_t|x_t,e_t,b_t,y^{t-1}) \nonumber\\
&& \IEEEeqnarraymulticol{1}{r}{\times \log \frac{u_t(y_t|x_t,e_t,b_t)}{p(y_t|y^{t-1})},} \nonumber \\
&=& \sum_{s_t \in \mathcal{S}, y_t \in \mathcal{Y}}  \beta_t(s_t) u_t(y_t|s_t) \log \frac{u_t(y_t|s_t)}{\sum_{\substack{\tilde{s}_t \in \mathcal{S}}} \beta_t(\tilde{s}_t) u_t(y_t|\tilde{s}_t)}, \nonumber \\
&=& I(X_t,E_t,B_t;Y_t|\beta_t,u_t),
\end{IEEEeqnarray}
where the last step confirms that $Y^{t-1} \rightarrow (\beta_t, u_t) \rightarrow (X_t,E_t,B_t,Y_t)$ is a Markov chain.

The following lemma summarizes the results of this section.

\begin{lemma}
Without loss of optimality, the SM privacy problem (\ref{eq:generalMinimization}) for the scenario in which the UP does not know the realizations of the renewable energy process can be modeled as an MDP, such that
\begin{enumerate}
\item the state at time $t$ is given by (\ref{eq:state}),
\item the action at time $t$ is specified by $u_t(y_t|x_t,b_t,e_t)$,
\item and the instantaneous cost is given by (\ref{eq:costUpdate}).
\end{enumerate}
\end{lemma}

In order to formulate the Bellman equations, it is convenient to first define an operator $T$ for the DP as follows
\begin{IEEEeqnarray}{c}
(TJ)(\beta) = g(s,\pi(\beta),\beta) + \nonumber \sum_{s\in \mathcal{S}, y \in \mathcal{Y}} \beta(s) u(y|s) J(\phi(\beta,u)),
\end{IEEEeqnarray}
for $\beta \in \bar{\mathcal{B}}$, where $J: \bar{\mathcal{B}} \rightarrow \mathbb{R}$ is the value function.

For the finite horizon setting, let $J_t$ denote the value function at time $t \leq n$, with $J_{n+1}=0$. For $t\leq n$, we have
\begin{equation}\label{eq:finiteHorizon}
J_{t}(\beta)=\inf_{u \in \mathcal{U}}[T J_{t+1}](\beta).
\end{equation}

The minimization problem is solved by going backwards in time from $t=n$ to $t=1$ in order to find the optimal policy $\pi^*=(\pi_1^*, \pi_2^*, \ldots, \pi_n^* )$ that minimizes (\ref{eq:finiteHorizon}) for every $t$.



In the infinite horizon scenario, since the total information leakage over an infinite number of stages is generally infinite, we minimize the average information leakage per stage, i.e.,
\begin{equation}\label{eq:averageCost}
J(\beta_0)=\lim_{n \rightarrow \infty} \frac{1}{n} \mathbb{E}^{\pi}\bigg[\sum_{t=0}^{n -1} g_t(s_t,\pi_t(\beta_t),\beta_t) \bigg].
\end{equation}

The solution for the infinite horizon problem can be determined as the solution to the following Bellman equation
\begin{equation}\label{eq:infiniteHorizon}
\lambda + J(\beta) = \inf_{u \in \mathcal{U}}[T J](\beta) ,
\end{equation}
where $\lambda \in \mathbb{R}$ is the optimal average information leakage, and the vector $J(\beta)$ is the relative or differential privacy leakage, i.e., the difference of the expected leakage to reach a conventional state and the cost that would be incurred if the cost per stage was equal to $\lambda$ for all states.
Via efficient dynamic programming algorithms, e.g., value iteration and policy iteration \cite{Bertsekas:2007}, (\ref{eq:infiniteHorizon})  can be solved and an optimal stationary policy $\pi^*=(\pi^*, \pi^*, \ldots, \pi^* )$ can be found.

\begin{proposition}\label{prop:valueFunction}
The value functions $\{J_t\}_{t=1}^{n}$ are concave.
\end{proposition}

Concave value functions allow the use of convex optimization algorithms. Finally, the following corollary generalizes our result to an input load and a RES with larger memory.
\begin{corollary}\label{cor:higherOrder}
Let $X$ and $E$ be Markov chains of order $m$ and $l$, respectively. Let $S_t \triangleq (X_{t-m+1}^t,E_{t-l+1}^t,B_t)$. The previous steps follow also for this scenario, where
\begin{enumerate}
\item the state is $\beta_t(s_t) \triangleq p(s_t|y^{t-1})$,
\item the action is $u_t(y_t|s_t)$,
\item and the cost is $g_t(s_t,u_t,y^{t}) \triangleq \log \frac{u_t(y_t|s_t)}{p(y_t|y^{t-1})}$.
\end{enumerate}
\end{corollary}

\section{Renewable Energy Known by the UP}\label{sec:EnergyKnown}
\begin{figure}[!t]
\centering
\includegraphics[width=1\columnwidth]{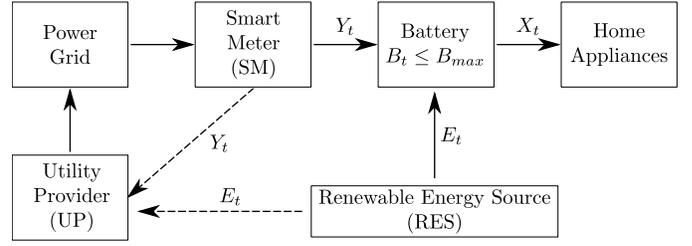}
\caption{System model. In this scenario, information about the realizations of the renewable energy process is available to the UP.}
\label{fig:SystemModelEnergyKnown}
\end{figure}
Here we assume that the UP knows the realizations of the renewable energy process $E^n$, as shown in Figure \ref{fig:SystemModelEnergyKnown}. This scenario can occur if we consider solar energy as the RES, and the UP can accurately estimate the renewable energy produced from its own observations in nearby locations, weather forecast of the area, and the specifications of the solar panel.

The goal is to find a battery charging policy $f\in \mathcal{F}$ that minimizes the following information leakage rate
\begin{IEEEeqnarray}{rCl}\label{eq:ppfE1}
\mathcal{I}(n,f ) &=& \frac{1}{n} I(X^n,E^n,B_1;Y^n|E^n), \nonumber\\
&=& \frac{1}{n} I(X^n,B_1;Y^n|E^n),
\end{IEEEeqnarray}
where the charging policy at time $t$ is
\begin{equation*}
f_t:\mathcal{X}^t \times \mathcal{E}^{t} \times \mathcal{B}^{t} \times \mathcal{Y}^{t-1} \rightarrow \mathcal{Y}, \qquad \forall t.
\end{equation*}

\subsection{Additive Formulation for the Information Leakage Rate}

Similarly to Section \ref{sec:EnergyNotKnown}, we want to express the problem in an additive form. We define policies $f'_t(y_t|x_t,e^{t},b_t,y^{t-1})$ as
\begin{equation}
f'_t:\mathcal{X} \times \mathcal{E}^t \times \mathcal{B} \times \mathcal{Y}^{t-1} \rightarrow \mathcal{Y}, \qquad \forall t,
\end{equation}
and state the following theorem.

\begin{theorem}\label{th:theoremppf2}
There is no loss of optimality in focusing only on charging strategies $f' \in \mathcal{F}'$, where $\mathcal{F}'\subseteq \mathcal{F}$. Moreover, the minimum information leakage rate can be written in the following additive form
\begin{equation}\label{eq:theoremppf2}
 \mathcal{I}(n) = \inf_{f' \in \mathcal{F}'} \frac{1}{n} \sum_{t=1}^{n} I^{f'}(X_t,B_t;E_t,Y_t|E^{t-1},Y^{t-1}).
\end{equation}
\end{theorem}

The proof follows similar steps to that of Theorem \ref{th:theoremppf}.

For the setting in which $X$ and $E$ are Markov processes of order $m$ and $l$, respectively, we define policies $f''$ such that
\begin{equation}\label{eq:policyGreaterOrder2}
f''_t:\mathcal{X}^m \times \mathcal{E}^{t} \times \mathcal{B} \times \mathcal{Y}^{t-1} \rightarrow \mathcal{Y}, \qquad \forall t.
\end{equation}
Then the following corollary holds.
\begin{corollary}\label{th:greaterOrderKnown}
Let $X$ and $E$ be Markov chains of order $m$ and $l$, respectively. There is no loss of optimality in focusing only on charging strategies $f'' \in \mathcal{F}''$, and for those strategies the minimum information leakage rate can be written in the following additive form
\begin{equation*}\label{eq:theoremppfGreaterOrderKnown}
 \mathcal{I}(n) = \inf_{f'' \in\mathcal{F}''} \frac{1}{n} \sum_{t=1}^{n} I^{f''}(X_{t-m+1}^t,B_t;E_t,Y_t|Y^{t-1},E^{t-1}).
\end{equation*}
\end{corollary}

\subsection{MDP Formulation}

As in Section \ref{subsec:MDP} we specify the state space, control actions and the instantaneous cost. The state of the DP at time $t$ is considered to be the causal posterior probability distribution over $S_t \triangleq (X_t,B_t)$, given the knowledge of $Y^{t-1}$ and $E^{t-1}$:
\begin{equation}\label{eq:stateCond}
\beta_t(s_t) \triangleq
\begin{cases}
p(s_1), & \text{if } t=1, \\
p(s_t|y^{t-1},e^{t-1}),       & \text{otherwise}.
\end{cases}
\end{equation}
$\beta_t(s_t)$ has again the interpretation of \emph{belief} that the UP has about $s_t$ at time $t$, given  $(Y^{t-1},E^{t-1})$.

The action $U_t \in \mathcal{U}$ is the conditional probability $u_t(y_t|s_t,e_t)$ given by $u_t=\pi_t(h)$. As before, we consider without loss of optimality Markov policies $\pi=\{\pi_1, \pi_2, \ldots\}$ that depend only on the current state $\beta_t$, i.e., $\pi_t(\beta_t)=\pi_t(h)$. As in (\ref{eq:recursion}), $\beta$ is updated recursively, i.e., $\beta_{t+1}=\phi'(\beta_t,u_t)$. $e_t$ is not in the belief as the UP has perfect knowledge about it.

We follow steps similar to those of Section \ref{subsec:MDP}, and define the cost of taking action $u_t$ when $s_t=(x_t,b_t)$ as
\begin{equation}\label{eq:instCostCond}
g_t(x_t,e^t,b_t,u_t,y^{t})\triangleq \log \frac{q_E(e_t|e_{t-1}) u_t(y_t|x_t,e_t,b_t)}{p(y_t,e_t|y^{t-1},e^{t-1})}.
\end{equation}

By considering the average $n$-horizon cost, it is possible to show that this formulation is equivalent to the original problem
\begin{IEEEeqnarray}{rcl}
  &\mathcal{I}&(n,\pi)= \frac{1}{n} \mathbb{E}^{\pi} \bigg[ \sum_{t=1}^{n}  g_t(x_t,e^t,b_t,u_t,y^t)\bigg], \nonumber \\
  &=& \frac{1}{n} \sum_{t=1}^{n} \sum_{\substack{s_t \in \mathcal{S} \\ y^t \in \mathcal{Y}^t, e^t \in \mathcal{E}^t}}   p(s_t,e_t,y_t|y^{t-1},e^{t-1}) g_t(s_t,e^t,u_t,y^{t}), \nonumber \\
  &=& \mathcal{I}(n,f'),
\end{IEEEeqnarray}
where $u_t$ is also a function of $(y^{t-1},e^{t-1})$ since $u_t=\pi_t(\beta_t)$. Given a policy $\pi$, $\beta_t$ and $u_t$ are determined by $(y^{t-1},e^{t-1})$. Then, it is possible to write
\begin{IEEEeqnarray}{rcl}\label{eq:costUpdateCond}
&\mathbb{E}&^{\mathbf{\pi}}\Big[g_t(X_t,E^t,B_t,U_t,Y^t)\Big] \nonumber \\
&=& I(X_t,B_t;E_t,Y_t|Y^{t-1}=y^{t-1},E^{t-1}=e^{t-1}),\nonumber \\
&=& \sum_{\substack{x_t \in \mathcal{X}, e^t \in \mathcal{E}^t \\ b_t \in \mathcal{B}, y^t \in \mathcal{Y}^t}}  p(x_t,e_t,b_t|y^{t-1},e^{t-1}) p(y_t|x_t,b_t,e^{t},y^{t-1}) \nonumber\\
&& \IEEEeqnarraymulticol{1}{r}{\times \log \frac{q_E(e_t|e_{t-1}) u_t(y_t|x_t,e_t,b_t)}{p(y_t,e_t|y^{t-1},e^{t-1})},} \nonumber \\
&=& \sum_{\substack{s_t \in \mathcal{S}, y_t \in \mathcal{Y}\\ e_t \in \mathcal{E}}}  \beta_t(s_t) q_E(e_t|e_{t-1}) u_t(y_t|s_t,e_t) \nonumber \\
&& \IEEEeqnarraymulticol{1}{r}{\times \log \frac{q_E(e_t|e_{t-1}) u_t(y_t|s_t,e_t)}{\sum_{\substack{\tilde{s}_t \in \mathcal{S}}} \beta_t(\tilde{s}_t) q_E(e_t|e_{t-1}) u_t(y_t|\tilde{s}_t,e_t)},} \nonumber \\
&=& I(X_t,B_t;E_t,Y_t|\beta_t,q_E,u_t),
\end{IEEEeqnarray}
where the last step confirms that $(Y^{t-1},E^{t-1}) \rightarrow (\beta_t, q_E, u_t) \rightarrow (X_t,E_t,B_t,Y_t)$ is a Markov chain.

The following lemma summarizes the results of this section.

\begin{lemma}
Without loss of optimality, the SM privacy problem (\ref{eq:generalMinimization}) for the scenario in which the UP knows the realizations of the process $E^n$ can be modeled as an MDP, such that
\begin{enumerate}
\item the state at time $t$ is given by (\ref{eq:stateCond}),
\item the action at time $t$ is specified by $u_t(y_t|x_t,e_t,b_t)$,
\item and the instantaneous cost is given by (\ref{eq:costUpdateCond}).
\end{enumerate}
\end{lemma}




Bellman equations for the finite and infinite horizon problems can be obtained as in Section \ref{sec:EnergyNotKnown}, with the consequent changes in the formulations of $\beta$, $u$ and $\pi$.

A final corollary, counterpart of Corollary \ref{cor:higherOrder}, holds.
\begin{corollary}
Let $X$ and $E$ be Markov chains of order $m$ and $l$, respectively, $S_t \triangleq (X_{t-m+1}^t,B_t)$, and $q_E^l$ the $l$-th order transition probability. The previous passages follow also for this scenario, where
\begin{enumerate}
\item the state is $\beta_t(s_t) \triangleq p(s_t|y^{t-1},e^{t-1})$,
\item the action is $u_t(y_t|s_t,e_t)$,
\item the cost is $g_t(s_t,u_t,y^{t},e^t) \triangleq \log \frac{q_E^l(e_t|e_{t-l}^{t-1}) u_t(y_t|s_t,e_t)}{p(y_t,e_t|y^{t-1},e^{t-1})}$.
\end{enumerate}
\end{corollary}

\section{Conclusions} \label{sec:conclusion}

We have studied the information leakage rate in an SM system by considering the availability of a RES and a finite capacity battery at the consumer side. The minimum information leakage rate has been characterized for both the scenario in which the UP does not know the realizations of the renewable energy process, and the scenario in which the UP knows them. For both scenarios, we have formulated the minimum information leakage rate as an additive cost function, and cast the problem as an MDP, thereby finding the expressions for the corresponding Bellman equations. The optimal leakage rate for a given scenario can be obtained by discretizing the continuous belief state and applying dynamic programming techniques.

\bibliographystyle{IEEEtran}
\bibliography{SPAWCref}

\end{document}